# Reconfigurable Intelligent Surface & Edge -- An Introduction of an EM manipulation structure on obstacles' edge

Tianqi Xiang
School of Information and Telecommunication Engineering
Beijing University of Posts and Telecommunications
Beijing, China
xiangtianqi@sina.com

Zhiwei Jiang
School of Information and Telecommunication Engineering
Beijing University of Posts and Telecommunications
Beijing, China
xiaojiangtongxue@bupt.edu.cn

Weijun Hong
School of Information and Telecommunication Engineering
Beijing University of Posts and Telecommunications
Beijing, China
hongwj@bupt.edu.cn

Xin Zhang
School of Information and Telecommunication Engineering
Beijing University of Posts and Telecommunications
Beijing, China
zhangxin@bupt.edu.cn

Yuehong Gao
School of Information and Telecommunication Engineering
Beijing University of Posts and Telecommunications
Beijing, China
yhgao@bupt.edu.cn

*Abstract*—Reconfigurable Intelligent Surface (RIS) or metasurface is one of the important enabling technologies in mobile cellular networks that can effectively enhance the signal coverage performance in obstructed regions, and it is generally deployed on surfaces different from obstacles to redirect electromagnetic (EM) waves by reflection, or covered on objects' surfaces to manipulate EM waves by refraction. In this paper, Reconfigurable Intelligent Surface & Edge (RISE) is proposed to extend RIS' abilities of reflection and refraction over surfaces to diffraction around obstacles' edge for better adaptation to specific coverage scenarios. Based on that, this paper analyzes the performance of several different deployment locations and EM manipulation structure designs for different coverage scenarios. Then a novel EM manipulation structure deployed at the obstacles' edge is proposed to achieve static EM environment modification. Simulations validate the preference of the schemes for different scenarios and the new structure achieves better coverage performance than other typical structures in the static scheme.

## I. Introduction

Looking beyond 5G mobile wireless communication system, large-scale MIMO [1], high frequency (mmWave as well as terahertz communication) [2] and reconfigurable intelligent surface (RIS) [3][4] are of great potential for improving coverage and capacity. Reconfigurable intelligent metasurface achieves manipulated transmission of electromagnetic (EM) waves by means of programmable amplitude and phase control units. This modifies the EM propagation environment to overcome adverse conditions, such as blockage, which is promising for mobile coverage enhancement.

However, most researches on RIS focus on propagation models [5][6] and theoretical performance [7], etc., where the communication link is highly abstracted and independent of any specified communication scenario. Although some researchers have conducted partial studies for specific scenarios [8][9], the choice of EM manipulation structures to be used and where to deploy them for different communication scenarios still need to be studied in detail. It is also worthwhile investigating whether state-of-the-art RIS structures, such as reflective/transmissive RIS [10], are enough for different deployment locations (e.g. building surfaces or edges) and what other types of EM manipulating structures can be introduced. Simultaneously transmitting and reflecting RIS (STAR-RIS) is designed for the purpose of full space coverage [11], and when it comes to region-specific enhancement, it is no different from reflective or transmissive RIS. Beyond diagonal RIS (BD-RIS) has been proposed to address the challenges of diverse coverage environments [12][13], which however, lacks practical manipulating structural design. It may also focus on full area coverage rather than some specific areas, such as shaded regions behind buildings. It should be noted that, these previous works have not considered or tried to exploit the effect of diffraction around obstacles' edges, and they have not investigated the EM manipulation structures composed of surface elements combing intentionally designed edge structures as proposed in this paper.

The contributions are listed below. An evaluation framework is built for assessing scenario-dependent coverage enhancement performance, while introducing the concept of Reconfigurable Intelligent Surface and Edge (RISE) to represent EM enhanced structures deployed at different locations, including edge structure (ES) and surface structure (SS). A novel edge EM manipulation structure, composed of an ultrathin perfect polarization rotator (UPPR) and a waveguide, named diffraction enhancement edge (DEE), is also proposed to specifically improve the ES scheme.

In section II, we introduce the evaluation framework for RISE, as well as the structure of DEE. Section III analyses several aspects of deployment geometry differences, overhead, cost and backward compatibility, and proposes a static enhancement scheme. In section IV, simulation results show that the ES achieves better results in building edge occlusion scenarios, and the potential of DEE is discussed. Section V and VI gives the future work and conclusion respectively.

## II. SYSTEM MODEL

### A. Geometry and Scenarios

In this paper, we consider the coverage of a typical cellular mobile communication system in the blocking region of a base station that is obstructed by a building as shown in Fig.1. It is assumed that there is another building wall plane besides the obstacle building to deploy SS and a building edge to deploy ES (which could be transmissive RIS at a certain angle or the proposed new structure, corresponding to the left ES and the right one in Fig. 1, respectively).

Two scenarios are considered to evaluate the performance of RISE, as shown in the Fig.2, where (a) indicates coverage enhancement of region blocked by a building corner, and (b) indicates coverage enhancement of region obscured behind a wall. In order to compare the coverage performance of the schemes, the pathloss distribution will be evaluated and analyzed under different scenarios and parameter settings. Fig.2 gives 2D layouts and coordinates in meters, the intended coverage areas are marked with shadow; the heights of the building, RIS, ES and TX/RX antenna are assumed to be the same.

### B. Pathloss Model

Considering a RIS link, let $d_1$ and $d_2$ be the distances from the BS to the RISE structure and from the structure to UE, $G_t$ be the BS antenna gain, $G_r$ be the UE antenna gain, and $G$ be the maximum gain of the metasurface structure. As illustrated in fig. 3, the incident direction is indicated by $(\theta_i, \varphi_i)$, and the observation direction is indicated by $(\theta_s, \varphi_s)$. When the typical reflective/transmissive metasurface [10] is considered, there are $M$ rows and $N$ columns of basic units. Each basic unit is $d_x$ in width, and $d_y$ in height. The incident pattern is expressed by $F_i(\theta_i, \varphi_i)$, and scattered pattern expressed by $F_s(\theta_s, \varphi_s)$. The magnitude phase excitation of each unit is $Ae^{j\phi_{m,n}}$. $0 < \Gamma \leq 1$ is the attenuation coefficient of reflection/transmission. Let $\lambda$ be the wavelength. Then the free space path loss of the typical reflective/transmissive metasurface can be expressed as [6]:

$$PL = \frac{\Gamma G_t G_r G M^2 N^2 d_x d_y \lambda^2 F_i(\theta_i,\varphi_i) F_s(\theta_s,\varphi_s) A^2}{64\pi^3 d_1^2 d_2^2} |\beta|^2 \quad (1)$$

where β is the normalized gain due to phase excitation [6]:

$$\beta = \frac{1}{MN} \sum_{n=1}^{N} \sum_{m=1}^{M} e^{j(\phi_{m,n}+\phi_{m,n}^i)} \quad (2)$$

where the incident wave phase at the $i$-th unit is expressed by [6]:

$$\phi_{m,n}^i = \mod\left(\frac{2\pi}{\lambda}\left((\sin\theta_i \cos\varphi_i + \sin\theta_s \cos\varphi_s)\left(m - \frac{1}{2}\right)d_x + (\sin\theta_i \sin\varphi_i + \sin\theta_s \sin\varphi_s)\left(n - \frac{1}{2}\right)d_y\right), 2\pi\right) \quad (3)$$

For the desired beam direction $(\theta_d, \varphi_d)$, the phase excitation of the $i$-th unit is calculated by [6]:

$$\phi_{m,n} = \mod\left(\frac{2\pi}{\lambda}\left((-\sin\theta_i \cos\varphi_i - \sin\theta_d \cos\varphi_d)\left(m - \frac{1}{2}\right)d_x + (-\sin\theta_i \sin\varphi_i - \sin\theta_d \sin\varphi_d)\left(n - \frac{1}{2}\right)d_y\right), 2\pi\right) \quad (4)$$

The incident wave pattern is expressed by [5][6]:

$$F_i(\theta_i, \varphi_i) = \cos^2 \theta_i \quad (5)$$

The scattered wave pattern is expressed by [6], where $\alpha$ denotes a fit parameter:

$$F_s(\theta_s, \varphi_s) = \cos^\alpha \theta_s \quad (6)$$

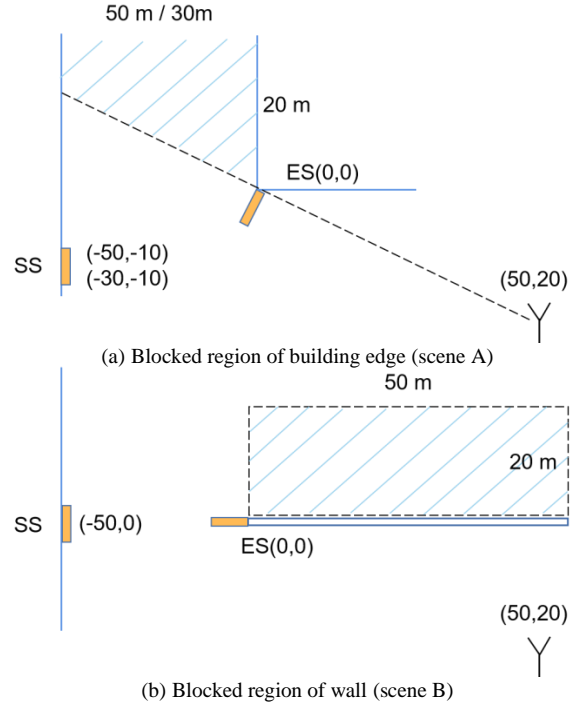

Fig. 2. Evaluation scenarios

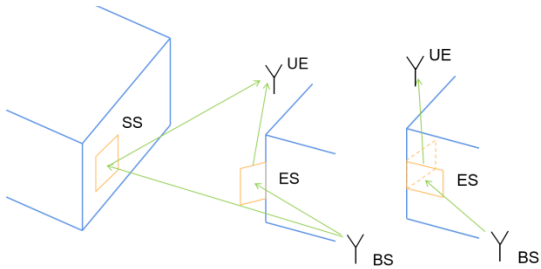

Fig. 1. The deployment of SS and ES

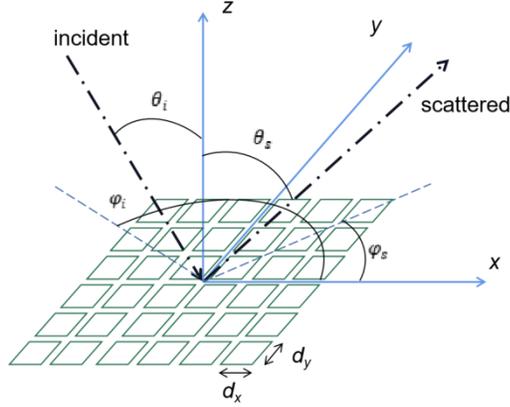

Fig. 3. Illustration of RIS

Equation (1) gives that the pathloss of the metasurface structure is influenced by the two path lengths $d_1$ and $d_1$, i.e. $PL \propto (d_1 d_2)^{-2}$. Here the effect of the two path lengths is multiplicative. As the unit plate area $d_x d_y$ and number of units $MN$ increase, there will be larger received energy. $\Gamma$, $F_i(\theta_i, \varphi_i)$ and $F_s(\theta_s, \varphi_s)$ are determined by the metasurface structure; in general, for an ideal reflective metasurface, $\Gamma = 1$, for a transmissive metasurface, $\Gamma < 1$. In addition, as each unit quantizes the phase change into several discrete values, quantization error could occur, thus reducing the beamforming gain.

### C. A Novel EM Manipulation Structure Mounted on Edge

This paper proposes a novel EM manipulation structure for signal enhancement in ES scheme. This structure is divided into two parts: the ultrathin perfect polarization rotator (UPPR) [14] and the waveguide, as shown in fig. 4(a) and fig. 4(b), respectively.

The UPPR structure, as shown in fig. 5(c), is constructed by combining two dipole resonant polarization selectors (DRPS) as shown in fig. 5(a) on both sides and a cross-shaped polarization converter (CSPC) in fig. 5(b) in the centre, which is sandwiched between two dielectric layers. The CSPC is rotated 45° with respect to the DRPS. Due to the electromagnetic coupling between parallel metallic structures, chiral metamaterials could provide strong polarization rotation power, suitable for efficient transmission of cross-polarized waves.

The high characteristic impedance of waveguide is easily matched with half-wave slit antenna so waveguide becomes an ideal transmission line for slit antenna feed. When a microwave signal is transmitted in the waveguide, an inductive current will be generated on the surface of the inner wall of the metal waveguide. Waveguide slit is open in the waveguide wall. When the slit cuts off the current on the waveguide wall, there will be current flow to the outer wall of the waveguide and at the same time excites electric field between the slit. The distribution of the electric field can be equated to the axis of slit surface current. The currents on the outer wall of the waveguide and the magnetic currents on the slits radiate electromagnetic waves into space.

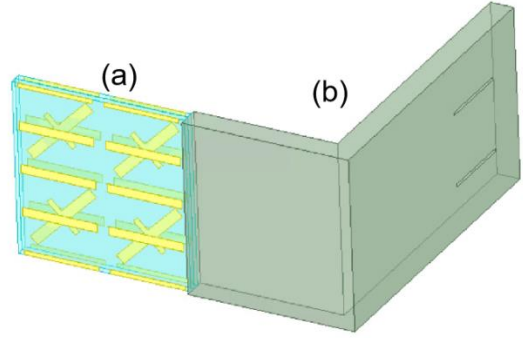

Fig. 4. The novel DEE structure mounted on edge

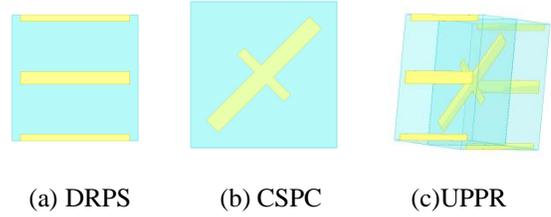

(a) DRPS  (b) CSPC  (c) UPPR

Fig. 5. The UPPR structure

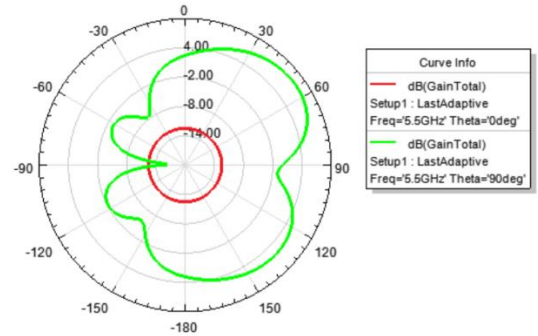

Fig. 6. The radiation pattern of DEE

Then the incident vertically polarized electromagnetic wave is fed into the waveguide through the UPPR that converts a portion of the vertically polarized energy into horizontally polarized energy, which is then radiated externally through the slits in the waveguide.

The new structure is called DEE here because its effect on signal propagation is related to the diffraction effect of the whole structure. When the signal propagates over the edge of an obstacle, the signal produces coverage of the shadow area behind the obstacle due to diffraction. The DEE improves the signal quality in the shadowed area behind the obstacle. Different from the reflection and refraction effect of typical RIS (including STAR-RIS and BD-RIS), DEE directs the signal around the edge to achieve enhancement specific to this particular shadowed scene.

As the complete physical model of the new structure's pathloss model still remains to be investigated, the following simplified model is proposed:

$$PL = \frac{G_t G_r \tilde{G} \lambda^2 \tilde{F}(\theta_i, \varphi_i, \theta_s, \varphi_s)}{16\pi^2 d_1^2 d_2^2} \quad (7)$$

where $\tilde{F}(\theta_i, \varphi_i, \theta_s, \varphi_s)$ is the normalized pattern from the incident direction $(\theta_i, \varphi_i)$ to the scattered direction $(\theta_i, \varphi_i)$; $\tilde{G}$ is the gain of the new structure. It should be noted that the unit of $\tilde{G}$ is m², which is positively correlated with the area of the structure. Here $\tilde{F}(\theta_i, \varphi_i, \theta_s, \varphi_s)$ and $\tilde{G}$ are derived with EM simulation software, such as HFSS. 45% radiation efficiency is achieved at 5 GHz-6 GHz, with the highest radiation efficiency of 62% at 5.5 GHz. Radiation pattern is obtained as in fig. 6.

## III. DIFFERENCE ANALYSIS BETWEEN SS AND ES

Due to the different deployment geometries in specific scenarios, the SS and ES differ in the following several perspectives.

### A. Propagation Path Length

For all the considered structures, the contribution to the final pathloss of the BS to RISE and RISE to UE pathloss is multiplicative. This effect leads to the signal attenuation being more sensitive to the propagation path length. In actual deployment scenarios, a reflective surface is not always suitable when it is hard to find a site to deploy it; even if such a site exists, the reflective surface tends to be further away than the ES deployment. Although ES is also subject to multiplicative fading, it tends to be located much closer than SS and the ES is less likely affected by this multiplicative fading effect.

### B. Angular Range of Coverage

SS achieves signal transmission by creating a new reflective path rather than dealing with the blockage directly. However, in this way the beam alignment becomes less dependent on the position of the blockage and more dependent on the position of users. When the blockage region behind the building is large, especially if the coverage zenith $\theta_s$ becomes extreme, it may result in lower gain in terms of radiation pattern. Even if the deployment angle of the SS can be adjusted, the simultaneous guarantee of a high incident pattern and scattered pattern is essentially a contradiction under the premise of reflective deployment. On the other hand, ES structures are inherently concerned with the angular range of coverage between the building edge and the user. Moreover, the rotation of ES could be a win-win case for both incident and scattered pattern gain increases. So the ES design is more suitable in terms of solving the obstacle blockage problem than the SS.

### C. Overhead

The overhead is determined by two main aspects: the number of RIS phase quantization bits and the number of units, both of which also determine whether the beam is aligned with good directivity and high gain. With a constant surface area, these two aspects determine the quantization accuracy of the phase and the phase difference between units respectively [5], which could lead to very high computational overhead when an excellent directivity is required for the desired beam. This effect may therefore limit the actual performance of the RISE.

TABLE I. RISE STRUCTURE PARAMETERS

| Parameters | Structures | | |
|---|---|---|---|
| | SS I | SS II | ES |
| $d_x$ | 0.01 m | 0.01 m | 0.01 m |
| $d_y$ | 0.01 m | 0.01 m | 0.01 m |
| M | 16 | 32 | 16 |
| N | 16 | 32 | 16 |
| $\Gamma$ | 1 | 1 | 0.5 |
| Quantization Level | 4 | 4 | 4 |
| α [6] | 3 | 3 | 3 |

### D. Deployment Costs

In order to obtain a higher energy density of the received signal, a large area or form factor is required for whatever structure is used, so that a high radiated energy density is possible. However, this can be very costly for RIS, as the basic unit is generally small in area with high cost of fine processing. The need for a large number of tiny units will increase the cost of RISE deployments. But the overall shapes of the structure for SS and ES would be quite different.

### E. Backward Compatibility Support

In general communication systems considering RISE, RISE channel estimation/RISE beam scanning is often necessary to ensure real-time gain, which introduces additional overhead along with modification to the communication protocol to support additional interactions. For channel sounding and beam scanning may consume too much overhead and require complex protocol support if each RISE structure participated in the real time beamforming, the use of fixed beams might be allowed, ensuring backward compatibility at the cost of losing some beamforming gain and flexibility.

More discussions are given based on simulation results in section IV.

## IV. SIMULATIONS AND ANALYSIS

Two sizes of Reflective and transmissive RIS are considered, as listed in table I. The frequency of wireless signal is 5.5GHz. Here reflective RIS is applied to SS, and we use ES to refer to the transmissive RIS at the edge. DEE denotes the structure proposed in this paper.

Simulations were carried out to obtain the pathloss in coverage regions of SS and ES for each scenario, which is shown in Fig. 7.

In building edge blocked regions, the ES's link has lower attenuation and the coverage is more consistent in this region, while poorer coverage can be easily observed for SS due to extreme scattered angles. This could also be supported by the results that the SS coverage becomes worse as the building spacing decreases, which leads to increased percentage of extreme angles. Conversely, the SS has a more consistent and low attenuation for wall-blocked environments. In this case, although the ES is closer to the blocked region but the extreme scattered angles also limit its performance.

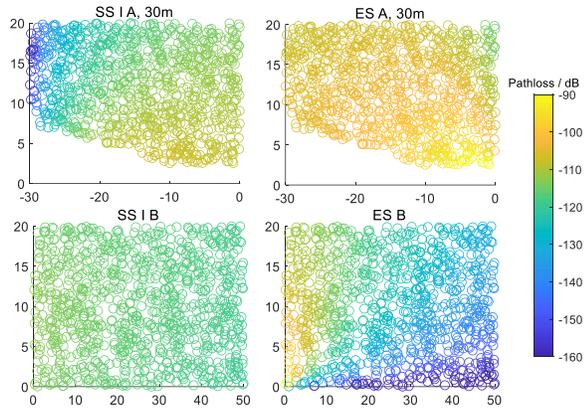

Fig. 7. Pathloss distribution map

Fig. 8 shows the cumulative distribution function (CDF) of the pathloss distribution for different structures in each scenario, where dynamic directional beams are considered in fig. 8(a) and a fixed beam in fig. 8(b). With the same configuration of metasurface, even though the transmissive metasurface suffers from penetration loss ($\Gamma = 0.5$), it still has low attenuation in scene A. Here we regard path length and angular coverage range as inherent geometrical differences between SS and ES enhancement schemes. As a trade-off between the two aspects for SS schemes, finding closer reflective surface deployment locations often requires greater angular coverage, and conversely, smaller angular coverage implies longer propagation distances. As shown in fig. 8(a), "SS I A, 50m" case suffers from higher attenuation due to larger path length, and on the contrary, "SS I A, 30m" case suffers from wider pathloss distribution because of larger angular range. But this contradiction does not exist for the ES scheme, which performs better than SS in scene A and the variation in performance over distance is slight.

And we see overhead and deployment costs as two complementary metrics of enhancement. For example, to increase gain, it is supposed to use a higher overhead configuration or increase the surface area. In general, link quality improvement requires both to be optimized collaboratively. And if a structure is more suitable for an environment, overhead and cost are correspondingly less required. As presented in fig. 8(a), a larger size SS with more units could obtain similar coverage performance as the ES scheme, indicating that the ES scheme suitable for scene A indeed reduces overheads as well as deployment costs. This could be attributed to the preference of different blocked regions for enhancement, with scene A preferring ES scheme and scene B preferring SS scheme.

In addition, the backward compatibility might be a vital consideration for building a communication-friendly environment. So a static configuration of RISE with fixed beam and pattern is simulated and analyzed. Fig. 8(b) shows the CDF of the path loss distribution when using a fixed beam for coverage, with each scenario producing a beam in only one direction. It can be seen that in each scenario there is a varying attenuation increase due to the deprivation of beam assignment. But here comes better backward compatibility as the upgrade of channel estimation and feedback associated with RISE is no longer required. In this case, RISE becomes a static enhancement structure that integrates into the environment. This might be an effective low overhead method of green communication. Furthermore, the ES scheme is still more suitable for scene A and SS is more suitable for scene B for scenario preference as mentioned earlier. The DEE achieves overall lower attenuation than other structures, making it as a better structure for the static ES scheme.

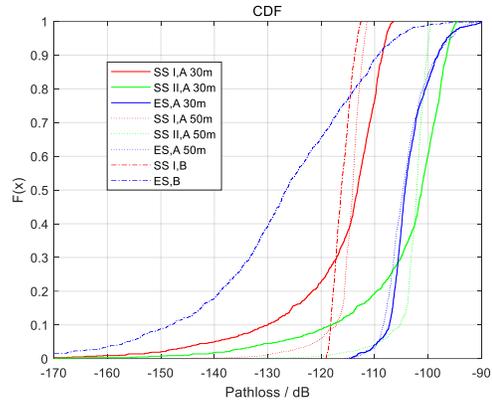

(a) Real-time beams

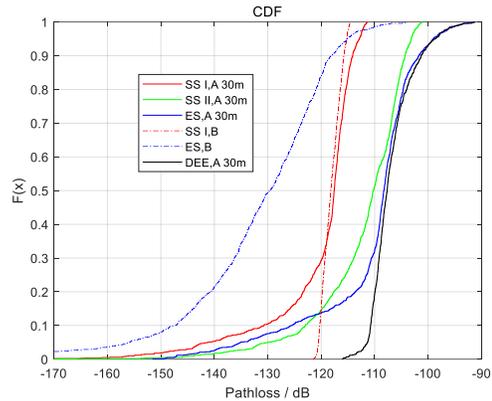

(b) Fixed beam

Fig. 8. Pathloss CDF

## V. Future Work

Hot research topics about RIS should also be applicable for ES individually, or for SS and ES jointly, i.e. RISE, such as channel estimation and optimization of the system, etc., besides, there are some other potential directions as follows.

Design of other new ES structures. At this stage, this new structure DEE achieves better coverage in static enhanced scenarios. However, in order to exploit its excellent properties, its manipulation, i.e., scattered/diffracted wave phase control could be further investigated. That is expected to dynamically obtain beamforming array gain, leading to better enhancement. Furthermore, it can be expected that, when the frequency reaches mmWave, THz or even much higher region, the form factor of the proposed RISE structure would be greatly reduced, and it might be possible to even design a string-shaped ES without SS, which would be much easier to deploy. However, the string-shaped ES requires more accurate beam alignment especially in dynamic scenarios, which may need some solutions to address this challenge as discussed below.

Combinations with other RF/non-RF sensing methods. In our previous work, for example, we have used computer vision to detect communication targets for low-overhead beamforming [15] and other beamforming-related applications [16]. Computer vision is able to not only detect communication targets, but also sense the environment, for example by extracting information about the edges and walls of buildings. Visual detection could be more instructive for beamforming when transmission is blocked, as after detecting building edges or surfaces, the beam can be aimed at RISE structures. Further, RISE structures could be specially marked for better identification. This approach is expected to significantly reduce the wireless overhead used for channel estimation and beam scanning, as well as enable efficient beam tracking.

## VI. CONCLUSION

Starting with the proposed RISE and scenario evaluation framework, this paper introduces geometry, scenarios and pathloss models, and proposes a new EM architecture mounted on the edge. Then the differences between ES and RIS in terms of geometric features, overhead, deployment cost and backward compatibility are analyzed. Simulation results show that ES is more suitable for building edge blocked region and RIS is more suitable for wall blocked region, while the new structure DEE performs better for static ES scheme.

## REFERENCES

[1] E. G. Larsson, O. Edfors, F. Tufvesson and T. L. Marzetta, "Massive MIMO for next generation wireless systems," in IEEE Communications Magazine, vol. 52, no. 2, pp. 186-195, February 2014.

[2] F. Akyildiz, C. Han and S. Nie, "Combating the distance problem in the millimeter wave and terahertz frequency bands", IEEE Commun. Mag., vol. 56, no. 6, pp. 102-108, Jun. 2018.

[3] Q. Wu and R. Zhang, "Towards Smart and Reconfigurable Environment: Intelligent Reflecting Surface Aided Wireless Network," in IEEE Communications Magazine, vol. 58, no. 1, pp. 106-112, January 2020.

[4] M. A. ElMossallamy et al., "Reconfigurable intelligent surfaces for wireless communications: Principles challenges and opportunities", IEEE Trans. Cognit. Commun. Netw., vol. 6, no. 3, pp. 990-1002, Sep. 2020.

[5] Ö. Özdogan, E. Björnson and E. G. Larsson, "Intelligent Reflecting Surfaces: Physics, Propagation, and Pathloss Modeling," in IEEE Wireless Communications Letters, vol. 9, no. 5, pp. 581-585, May 2020.

[6] W. Tang et al., "Wireless Communications With Reconfigurable Intelligent Surface: Path Loss Modeling and Experimental Measurement," in IEEE Transactions on Wireless Communications, vol. 20, no. 1, pp. 421-439, Jan. 2021.

[7] S. Zeng, H. Zhang, B. Di, Z. Han and L. Song, "Reconfigurable Intelligent Surface (RIS) Assisted Wireless Coverage Extension: RIS Orientation and Location Optimization," in IEEE Communications Letters, vol. 25, no. 1, pp. 269-273, Jan. 2021.

[8] Y. Li et al., "Path Loss Modeling for the RIS-Assisted Channel in a Corridor Scenario in mmWave Bands," 2022 IEEE Globecom Workshops (GC Wkshps), Rio de Janeiro, Brazil, 2022, pp. 1478-1483.

[9] M. Issa and H. Artail, "Using Reflective Intelligent Surfaces for Indoor Scenarios: Channel Modeling and RIS Placement," 2021 17th International Conference on Wireless and Mobile Computing, Networking and Communications (WiMob), Bologna, Italy, 2021, pp. 277-282.

[10] S. Zeng et al., "Reconfigurable Intelligent Surfaces in 6G: Reflective, Transmissive, or Both?," in IEEE Communications Letters, vol. 25, no. 6, pp. 2063-2067, June 2021.

[11] Y. Pan, Z. Qin, J. -B. Wang, Y. Chen, H. Yu and A. Tang, "Joint Deployment and Beamforming Design for STAR-RIS Aided Communication," in IEEE Communications Letters, doi: 10.1109/LCOMM.2023.3316620.

[12] Q. Li et al., "Reconfigurable Intelligent Surfaces Relying on Non-Diagonal Phase Shift Matrices," in IEEE Transactions on Vehicular Technology, vol. 71, no. 6, pp. 6367-6383, June 2022, doi: 10.1109/TVT.2022.3160364.

[13] H. Li, S. Shen and B. Clerckx, "Beyond Diagonal Reconfigurable Intelligent Surfaces: A Multi-Sector Mode Enabling Highly Directional Full-Space Wireless Coverage," in IEEE Journal on Selected Areas in Communications, vol. 41, no. 8, pp. 2446-2460, Aug. 2023, doi: 10.1109/JSAC.2023.3288251.

[14] Hong, W.; Zhu, J.; Deng, L.; Wang, L.; Li, S. Perfect terahertz-wave polarization rotator using dual Fabry–Pérot-like cavity resonance metamaterial. Appl. Phys. A 2019, 125, 1–7.

[15] T. Xiang, J. Gu, Y. Wang, Y. Gao and X. Zhang, "Computer Vision Aided Beamforming Fused with Limited Feedback," in *IEEE Transactions on Machine Learning in Communications and Networking*, doi: 10.1109/TMLCN.2023.3326409.

[16] T. Xiang, J. Gu, B. Yu and X. Zhang, "A Sensing-Based Over-the-Air Phased Array Antennas Phase Calibration Method With Greedy Algorithm," in IEEE Antennas and Wireless Propagation Letters, vol. 20, no. 12, pp. 2240-2244, Dec. 2021.